**Possible nanophotonics applications of the $V_N N_B$ defect in hexagonal boron nitride**

A. Sajid,[1,2#] Jeffrey R. Reimers,[1,3*] Rika Kobayashi,[3,4] and Michael J. Ford[1,3*]

[1] *University of Technology Sydney, School of Mathematical and Physical Sciences, Ultimo, New South Wales 2007, Australia.*
[2] *Department of Physics, GC University Faisalabad, Allama Iqbal Road, 38000 Faisalabad, Pakistan.*
[3] *International Centre for Quantum and Molecular Structures and Department of Physics, Shanghai University, Shanghai 200444, China.*
[4] *Australian National University Supercomputer Facility, Leonard Huxley Bldg 56, Mills Rd, Canberra, ACT, 2601, Australia.*

* Email: Mike.Ford@uts.edu.au, Jeffrey.Reimers@uts.edu.au
# current address: CAMD, Department of Physics, Technical University of Denmark, 2800 Kgs. Lyngby, Denmark

**ABSTRACT:** The $V_N N_B$ defect in hexagonal boron nitride (h-BN), comprising a nitrogen vacancy adjacent to a nitrogen-for-boron substitution, is modelled in regard to its possible usefulness in a nanophotonics device. The modelling is done on both a simple model compound and on a 2D periodic representation of the defect, considering its magnetic and spectroscopic properties. The electronic distribution in $V_N N_B$ excited states is very open-shell in nature, and to deal with this two new computational methods are developed: one allows standard density-functional theory (DFT) calculations to be employed to evaluate state energies, the other introduces techniques needed to apply the VASP computational package to these and many other problems involving excited states. Also of general use, results from DFT calculations are then calibrated against those from *ab initio* methods, seeking robust computational schemes. These innovations allow 45 electronic states of the defect in its neutral, +1 and -1 charged forms to be considered. The charged forms of the defect are predicted to display properties of potential interest to nanophotonics.

.

I.  INTRODUCTION

Point defects play an important role in spin and photophysics of semiconductor materials and can be exploited for technological application [1, 2]. Recently, 2D materials have emerged as a new class of semiconductors with color centers that have applications in many areas of nanophotonics [3-5], for example quantum sensing [6-9] and quantum information processing [10, 11]. Exploitation of these color centers requires a detailed knowledge of their electronic structure and magnetic properties. In particular, we focus herein on defects in hexagonal boron nitride (h-BN). Over the past few years, they have generated interest owing to the observation of single photon emission in both the visible [11-20] and UV spectral regions [21-23], with recent advances including the discovery of optically detected magnetic resonance [24, 25] (ODMR) and the identification of carbon as a constituent in many active defects [26]. Computational modelling of magnetic and optical properties has been critical to the establishment of assignments of observed signals to defects of particular chemical compositions [24, 26-33], but nevertheless many observed features of the significant visible emission observed in the 1.6 – 2.2 eV range still await interpretation [29]. The search for improved computational methodologies to facilitate this drives many aspects of modern research [34-36], and we look forward to the day in which defects of prescribed composition can be made, and assembled to order, based on predicted nanophotonics properties. Both of these aspects are addressed herein, considering the possible usefulness of the $V_NN_B$ defect (nitrogen vacancy with adjacent nitrogen-for-boron substitution) in h-BN, and in doing so developing new computational methodologies of general usefulness throughout defect spectroscopy.

Various studies have focused on identifying the nature of observed defects in h-BN [24, 26-34, 37-41], often inspired by the suggestion [28, 37] that defects such as $V_NN_B$, $V_NC_B$ and $V_BC_N$ could be implicated. Recent work has focused on the possibility that out of plane distortion [34, 42] control defect spectroscopic properties [33]. Studies of $V_BC_N$ revealed that computational methods normally thought to be reliable can show catastrophic failure when used in defect modelling, as most defect states are open-shell in nature, a feature poorly supported by most methods [26, 34]. Further, electronic transitions in defects can involve charge transfer, a feature that can also induce catastrophic failure in computations, demonstrated in an extreme example recently for the $V_N^-$ defect [36].

Very large spatial extent, coupled with very complex electronic interactions, makes accurate spectral predictions for h-BN defects extremely difficult. Multi-length-scale approaches such as mixed quantum-mechanics/molecular mechanics (QM/MM) offer a generally useful technologies for the future, allowing long-range nuclear structure to be modelled at a simple level, whilst critical short-range electronic interactions are treated at the highest level possible [33]. Such calculations can include multiple h-BN layers and large numbers (eg., 30) of rings of BN atoms surrounding the defect centre. Various modern calculations report the critical defect electronic properties as converging very quickly with model size expansion, with small 1-ring models being qualitatively reasonable but not always quantitatively accurate, 2-ring models adequate for most purposes, and 3-ring models showing near complete convergence [26, 33, 35, 36]. Defect models include both the cluster-type molecular models just described, as well as 2D (or 3D) models of periodic sheets (or solids). Even though 2D models must be much larger than cluster models to achieve convergence to similar accuracy for spectroscopic transitions confined within the defect, both types of approached lead to the same converged answers [35].

One feature favoring the efficient modelling of defects in 2D materials such as h-BN is the short-range nature of 2D dielectric effects compared to the long-range effects that are well-known to dominate 3D materials [43]; this facilitates the use of small model systems in electronic-structure calculations. Alternatively, defect features can make modelling very computationally challenging. Relative defect state energies may have very little correlation with differences in 1-particle orbital energies [34, 35], inhibiting the use of human intuition for state searching. Also, very large reorganization energies of several eV can be associated with geometrical relaxation following state transition, making the lowest-energy states often very difficult to determine [26, 33, 34].

The defect considered in this work, $V_NN_B$, is chosen as it has been widely mooted as a useful defect in nanophotonic applications [11, 12, 28, 30, 44], has electronic properties known to converge very rapidly with model size [35], yet presents some of the most difficult challenges for computational methods in terms of dealing with open-shell state character. These issues are so extreme that computational packages like VASP [45, 46], and others that evaluate spectroscopic properties through state-energy differences, are not even able to make estimates for many of its states. Beyond that, the VASP code in particular does not have the resources needed to converge the required electron densities needed to evaluate the state energies. Our first task is therefore to develop the required infrastructure, something useful throughout the field of defect research.

Our second task is then to calibrate the medium-level density-functional theory (DFT) methods available in many codes such as VASP against high-level density-functional and *ab initio* methods available in codes such as Gaussian-16 [47] and MOLPRO [48]. This work focuses on a simple 1-ring model compound, as well as a basic 2D periodic model of the defect. Form it, useful estimates are made of the properties of 45 electronic states of the defect in its neutral, +1 and -1 charged forms. This allows the desirability of the construction of $V_NN_B$ defects to be assessed.

## II. COMPUTATIONAL METHODS

Calculations were performed for periodically replicated defects in 2-D h-BN using DFT, as well as for a 1-ring model compound using both DFT and *ab initio* approaches. Many calculations refer to vertical excitation energies obtained at the ground-state structures shown in Fig. 1. Others refer to relaxed structures, yielding adiabatic transition energies and reorganization energies. All structures are constrained to local $C_{2v}$ point-group symmetry, and all symmetry specifications are named using standard conventions for planar molecules [49, 50], with the in-plane $a_1$ and $b_2$ axes shown in the figure.

For the 2D simulations, calculations of the total energies, electronic structures and optimized geometries were obtained using version 5.3.3 of VASP [45, 46]. For accurate calculation of electron spin density close to the nuclei, the projector augmented wave (PAW) method [51, 52] was applied, together with a plane-wave basis set. We utilized the standard PAW-projectors provided by the VASP package using a plane-wave basis-set cut-off of 350 eV. A large vacuum region of 30 Å width was used to separate a single layer of h-BN from its periodic images to minimize interactions. The defect was then realized in a fixed-size rectangular (6 x 4√3)R30° supercell (see Fig. 1b), with the native h-BN B-N bond length set to 1.443376 Å, treated only at the gamma-point of the Brillouin zone. Convergence of calculations for $V_NN_B$ with respect to sample size are rapid and have been considered in detail elsewhere, with likely shortcomings in the current calculations being in the order of just 0.01 eV [35]. Geometries are allowed to relax until a maximum atomic force of 0.01 eVÅ$^{-1}$ was reached. We used the PBE [53] and HSE06 [54, 55] density functionals to approximate electron exchange and correlation. All HSE06 2D optimized geometries are reported in Supplementary Material (SM) Section S5.

For the evaluation of magnetic properties, optimized 2D geometries were used to calculate the wave function coefficients of defect orbitals. Zero-field splitting tensors were evaluated using the method described by Ivády et al. [56], with full details of the calculations of spin properties described in SM Section S3. For the calculation of the spin-spin contribution to the zero-field tensor, a higher cut-off energy of 600 eV and lower force tolerance of $10^{-4}$ eV Å$^{-1}$ were used. The Hessian matrix specifying the normal modes depicting phonon motions was calculated using an energy cut off of 500 eV.

The model molecular compound used is shown in Fig. 1a. This contains a single ring of atoms surrounding the defect vacancy, augmented by three additional atoms near the nitrogen substituent. Additional the extra atoms produces a more balanced description of the defect's electronic properties [34], enabling very rapid convergence of electronic-state properties with respect to increasing ring numbers [35]. For spectroscopic transitions localized to the defect, the obtained results are estimate to be within 0.1 eV of values converged with regard to sample size [35]. In the vertical excitation-energy calculations initially performed, the cluster geometry used was obtained by extracting atoms from the 2D model (Fig. 1a), adding terminating hydrogen atoms at coordinates optimized using HSE06/6-31G* using Gaussian-16 [47].

A wide variety of calculations are performed on the model compound. DFT calculations using the HSE06 [54, 55] and CAM-B3LYP [57-59] density functionals using Gaussian-16 [47], with additional calculations in a large period cell of dimension $13 \times 30 \times 30$ Å performed using VASP [45, 46]. Time-dependent DFT (TD-DFT) calculations [60] are performed using Gaussian-16. Also applied are *ab initio* methods: coupled-cluster singles and doubles (CCSD) [61-64], this perturbatively corrected for triples excitations, CCSD(T) [65], complete active space self-consistent field (CASSCF) [66-68] calculations corrected for singles and doubles excitations, either perturbatively (CASPT2) [69], or else through contracted [70] multi-reference configuration (MRCI) [71] with Davidson correction, and equation-of-motion coupled cluster singles and doubles (EOM-CCSD) [72, 73] theory. The CCSD, CCSD(T), CASPT2, and MRCI calculations are performed using MOLPRO [74] and the EOM-CCSD calculations using Gaussian-16 [47]. Convergence of the CASPT2 and MRCI calculations with respect to choice of active space is addresses in SM Table S1. Comparison of best-available results to those obtained using smaller active spaces indicate convergence of results to within 0.13 eV for MRCI and 0.15 eV for CASPT2.

## 2. RESULTS AND DISCUSSION

### A. Overview

All calculations report that the ground state of neutral $V_NN_B$ (Fig. 1) as $(1)^2B_1$, when the defect is constrained to $C_{2v}$ symmetry [16, 37, 39]. Its ground state has been predicted to distort out of plane [42], but the consequences of this effect (which in defects can be critical [33]) are not significant to our interests herein, and its inclusion would make difficult the exhaustive scans that we perform of state manifolds. All work presented herein thus pertains to planar structures only. The basic electronic-orbital structure of $V_NN_B$ has also been discussed elsewhere [39], with critical features portrayed in Fig. 2. At each defect site, one σ and one π orbital produce dangling bonds. For the case of neutral $V_NN_B$, five electrons need be distributed in these six orbitals, reducing to four electrons for $V_NN_B^{+1}$, but increasing to six electrons for $V_NN_B^{-1}$. The three σ atomic defect orbitals combine to make molecular orbitals, one of an apparently "bonding" nature (named $1a_1$), one of a "non-bonding" nature (named $2a_1$), and one of an "antibonding" nature (named $b_2$). Similarly, the three π orbitals combine to make the orbitals named $1b_1$, $2b_1$, and $a_2$, respectively. The "bonding" orbitals are typical of 3-center 2-electron bonds but the interatomic distances are so large that in reality no bond exists (Fig. 1), and it is this open-shell feature that computational methods based on single-determinant representations of the wavefunction or density find difficult to model [34, 75]. We find that $1a_1$ lies below the valence-band (VB) maximum of h-BN and so is always doubly occupied, $2a_1$, $1b_1$ and then $2b_1$, shown pictorially in Fig. 2, fall in the VB to conduction band (CB) gap becoming the primary focus of attention, while $a_2$ and $b_2$ fall inside the CB. Defect orbital energies from spin-unrestricted HSE06 calculations on the 2D periodic model for neutral $V_NN_B$ are shown in the figure as a guide.

Transitions amongst the shown defect orbitals are expected to dominate the low-energy spectroscopy of the defect, with contributions also from VB and CB orbitals. We consider up to 18 electronic states for neutral $V_NN_B$, 14 for $V_NN_B^{+1}$, and 13 for $V_NN_B^{-1}$. These states are depicted in Table I in terms of their dominant orbital occupancies, with occupancy variations found involving 4 of the 6 defect orbitals, 6 VB orbitals, and 8 CB orbitals. Details of the VB and CB orbitals are model-size dependent, so the result obtained using the molecular model for states involving their partial occupation are only indicative.

Of the 45 states listed in Table I, 28 have dominant occupancies in which the number of partially occupied orbitals corresponds to that expected for the spin multiplicity (0 for a singlet state, 1 for a doublet state, 2 for a triplet state, of 3 for a quartet state). For many of these, single determinants can be used to represent the wavefunction, and the Kohn-Sham [76] or Gunnarsson-Lundqvist [77] theorems, etc., indicates that single-reference methods such as orthodox DFT and CCSD can be used to determine the state energies, with transition energies therefore determinable as state-energy differences. The theorems limit applications in that they indicate that results can only be obtained for the lowest-energy state of each spin multiplicity and spatial symmetry. Nevertheless, most computational packages allow higher roots to be obtained by setting the desired occupancies and using poor convergence criteria; results obtained in this fashion always require detailed scrutiny.

In addition to these states, 17 states listed in Table I have lower spin multiplicities than the maximum possible for the given number of unpaired electrons. For these open-shell states, multiple determinants must always be used to represent the wavefunction, an effect known as static electron correlation. The Kohn-Sham [76] and Gunnarsson-Lundqvist [77] theorems do not apply, so in principle DFT cannot be used to predict properties.

B. DFT calculations on open-shell systems.

We focus on the 17 states listed in Table I with essential open-shell character. If the ground-state of any system is essentially singlet reference, then excited states of any nature may be prepared using the first-principles TD-DFT approach (or EOM-CCSD based on a CCSD ground-state wavefunction). Alternatively, if adding or removing one or two electrons results in a closed-shell reference state, then first-principles approaches are also available [26, 78-83], something highly appropriate for the singlet states of $V_B^-$ [26]. In *ab initio* approaches, multi-reference methods like CASSCF, CASPT2, and MRCI readily deal with the issues generated. Currently, much research effort is being spent in the development of first-principles DFT approaches that capture the same essential elements [84-92]; similar empirical approaches have been applied to $V_B^-$ [31], with mixed success [26]. These approaches all involve methodologies not available in most DFT computational packages, however, and herein we consider empirical schemes that make calculations on this type of state widely available.

Empirical schemes have for a long time been applied to the simplest case of interest, when two unpaired electrons combine to make a singlet state. This situation is depicted in Fig. 3b and is very common, arising, e.g., whenever optical excitation occurs from a closed-shell ground state of a molecule or material. It is also commonly produced as a result of antiferromagnetic couplings in molecules and materials. As Fig. 3b shows, four determinants can be constructed depicting one singlet state and one triplet state. Determinants 5a and 5b each have two electrons of the same spin and so provide easy computational access to the properties of the triplet state. In contrast, determinants 6a and 6b, with one electron of each spin, do not satisfy the Gunnarsson-Lundqvist theorem and must be coupled together to make singlet-state and triplet-state components. Ignoring dynamic electron correlation between these electrons and the others, the energy of the mixed-spin determinants must be the average of the energies of the triplet and singlet states:

$$E_S + E_T = E_{6a} + E_{6b} \tag{1}$$

where $E_S$ and $E_T$ are the observable energies of the singlet and triplet states, while $E_{6a}$ and $E_{6b}$ are the energies coming from DFT calculations made by setting occupancies to correspond to each individual determinant. As $E_{6a} = E_{6b}$ in the absence of a magnetic field, and as the Gunnarsson-Lundqvist theorem yields $E_T = E_{5a} = E_{5b}$, this allows the singlet state energy to be approximated by

$$E_S = 2E_{6a} - E_{5a} . \tag{2}$$

Such an approach could be thought of as being the simplest-possible empirical multi-configurational DFT scheme. When considering antiferromagnetic interactions in materials, the electrons being correlated are typically large distances apart, making dynamical electron correlation small, and this approximation has successfully led to very many important results in the field. In defects, the neglect of dynamic electron correlation is questionable, but we have found that in $V_NC_B$ model compounds, the associated errors are small compared to the accuracy required to make qualitative assignment of defect properties [34]. Gradients for geometry optimizations should be taken from $E_S$, but instead we simply optimize $E_{6a}$ and then apply the correction from Eqn. (2), as the computations are much more likely to complete successfully.

Paralleling this well-known case for two electrons in two orbitals, we develop a similar empirical scheme for the case of three electrons in three orbitals. For neutral $V_NN_B$, the states $(2)^2A_1$, $(3)^2A_1$, $(3)^2A_2$ and $(5)^2A_2$ listed in Table I involve three unpaired electrons in three orbitals, with the spin multiplicity (doublet) being less than the highest-possible value (quartet). For each of these, 8 determinants contribute to the various states involved, as sketched in Fig. 3a. Four determinants, named 1a-4a, have more spin-up electrons than spin-down ones, while the others named 1b-4b, are symmetrically equivalent determinants obtained by interchanging spin-up and spin-down electrons. Ignoring spin-orbit coupling, in the absence of an applied magnetic field, the two sets of determinants have equal energies and are non-interacting, meaning that just one set, here taken as 1a-4a, need be explicitly considered in most discussions. The 8 determinants depict 3 electronic states: one quartet state with four spin components, and two doublet states each with two spin components. These are known as the *tripdoublet* state, labelled "D3" in Table I, and the *singdoublet* state, labelled "D1"; these names reflect the asymptotic limits of the wavefunction forms [93, 94]. Within a state, all spin components have the same energy. Two components 1a and 1b of the quartet state are immediately apparent in Fig. 3a, but the other two components, and all components of the doublet states, arise from complex linear combinations of determinants 2a-4a and 2b-4b. A first-principles multi-state DFT approach to deal with tripdoublet and singdoublet states has recently been developed [91, 95], but this is not available in most periodic codes widely applied by the h-BN community.

Determining an empirical scheme for the three-electrons in three-orbitals case is more difficult owing to the increased number of independent variables that need to be determined. As before, one equation of constraint can be written that conserves the total energy (trace of the Hamiltonian matrix) for the states:

$$E_{D1} + E_{D3} + E_Q = E_{2a} + E_{3a} + E_{4a} = E_{2b} + E_{3b} + E_{4b} ,  \quad (3)$$

where $E_{D1}$, $E_{D3}$, and $E_Q$ are the physically observable energies of the singdoublet, tripdoublet, and quartet states, respectively, in the absence of a magnetic field. As the Gunnarsson-Lundqvist theorem yields $E_Q = E_{1a} = E_{1b}$, this allows the average energy of the two doublet state to be approximated by

$$E_{D1} + E_{D3} = E_{2a} + E_{3a} + E_{4a} - E_{1a} . \quad (4)$$

Performing standard DFT calculations on the single determinants in Fig. 3a therefore cannot reveal the energies of the two doublet states, only their average. TD-DFT calculations explicitly embody all terms in a first-principles way and hence do not suffer from this problem, but they are currently not feasible to apply to large periodic solids containing defects. Hence we proceed by evaluating the energy splitting $\Delta E_{D1D3}$ using TD-DFT calculations for the model compound. The state energies are then approximated by:

$$E_{D3} = (E_{2a} + E_{3a} + E_{4a} - E_{1a} - \Delta E_{D1D3})/2,$$
$$E_{D1} = (E_{2a} + E_{3a} + E_{4a} - E_{1a} + \Delta E_{D1D3})/2. \quad (5)$$

In SM, Table S2 shows all of the quantities required to evaluate the energies of tripdoublet and singdoublet state pairs $(2)^2A_1$ and $(3)^2A_1$ (singly occupied orbitals $1a_1$, $1b_1$, and $2b_1$), as well as for the pairs $(3)^2A_2$ and $(5)^2A_2$ (singly occupied orbitals $V_{1a2}$, $1b_1$, and $2b_1$). The approximation

$$E'_{D3} \approx \min(E_{2a}, E_{3a}, E_{4a}) \quad (6)$$

provides an uncontrolled upper bound to the state energy. It has previously been applied to consider properties of $(2)^2A_1$ for the 2-D material using HSE06 [30], yielding results close enough to those from Eqn. (5) that are hence useful for assigning observed defect spectroscopy. Such an approach should be applied with extreme caution, however.

Gradients for geometry optimization should be taken from $E_{D3}$ or $E_{D1}$. Here, we take the simpler approach of optimizing the three individual configuration energies $E_{2a}$, $E_{3a}$, and $E_{4a}$, applying Eqn. (5) after the optimizations complete. This yields three different approximations for the state energies $E_{D3}$ or $E_{D1}$ and their associated optimized geometries. We find that the different results are all in good agreement, suggesting that this approach is successful.

C. DFT calculations using VASP that select states of prescribed spatial symmetry.

In SM Section S2, software is developed [30, 34, 96-99] that facilitates the use of the VASP package [45, 46] in determining defect state energies and densities. VASP does not provide the ability to determine the spatial symmetry of the wavefunctions that it optimizes. Software is provided that seamlessly determines the symmetry from the final listed wavefunction. Also, VASP allows initial occupancies to be selected, defining wavefunction symmetry, but does not guarantee that the final converged wavefunction will retain this symmetry. Software is described that overcomes this limitation by exploiting an available option that only partially optimizes the wavefunction, but most of the time retains symmetry. Through an iterative procedure, a fully optimized wavefunction, obeying the Gunnarsson-Lundqvist [77] theorem, but constrained to be of the desired symmetry, can usually be determined.

### D. *Ab Initio* and DFT calculations for the neutral $V_NN_B$ model compound test geometry.

The relative energies to the ground state, obtained using 11 computational methods for 13 low energy excited states, of the model compound (Fig. 1a) are compared in Table II, evaluated at the test geometry. The included states were chosen based on TD-DFT and EOM-CCSD evaluations, selecting all low-energy states plus also a few others of particular relevance. This systematic search procedure identified some low-energy states such as $(2)^2B_1$ and $(1)^2A_2$ that have not previously been considered [30].

First, we note that parallel calculations are reported using Gaussian-16 and VASP. Agreement of these calculations, performed using very different numerical approaches, is usually very good, with state energy differences reported as within 0.03 eV [35] or 0.08 eV [34] for states for which VASP has no difficulty in providing results. Table II shows mostly similar results, indicating the stability of the new numerical techniques introduced to allow VASP to function more generally; the exception is for $(2)^2A_2$, differing by 0.27 eV, and is indicative of the problems associated with operations on high-energy states of a given spin and spatial symmetry outside the bounds allowed by the Gunnarsson-Lundqvist [77] theorem.

To compare and calibrate methods, Table III lists the average differences and standard-deviations between excited state energies obtained by comparing the various computational results presented in Table II to each other. First we consider comparisons between results obtained using the *ab initio* MRCI, CASPT2, CCSD, EOM-CCSD and CCSD(T) methods. Each method has its own set of advantages and disadvantages in terms of feasibility and comprehensiveness. MRCI treats static electron correlation the best, with CASPT2 providing a more computationally efficient approximation, CCSD(T) includes static electron correlation in an asymmetric fashion yet contains the best description of dynamic electron correlation, provided that the occupied-unoccupied orbital energy difference is large enough, while CCSD is a more computationally efficient alternative that, in this application, is most likely more approximate than MRCI. EOM-CCSD embodies any deficiency in the treatment of its reference state (here $(1)^2B_1$) and has a less accurate treatment of dynamic electron correlation, but, like TD-DFT, properly includes the static electron-correlation effects addressed empirically through Eqns. (1)-(5).

The results show excellent agreement between the best methods, MRCI and CCSD(T), predicting an average excited-state energy difference of $0.0 \pm 0.3$ eV. The CCSD and EOM-CCSD results differ from these by 0.3 eV, suggesting that the enhanced treatment of dynamic correlation present in CCSD(T) is able to subsume the enhanced treatments of static correlation present in MRCI, with other correlation effects not being important. That CASPT2 agrees well with MRCI and CCSD(T) results supports this conclusion, suggesting that it provides an efficient and accurate *ab initio* approach for studying the spectroscopy of this defect. This situation is very different to that found for $V_NC_B$ [34] and $V_B^-$ [26]. Its singlet ground state leads to dramatic effects associated with static electron correlation, with reduced internal consistency between the different *ab initio* results.

Next we compare these *ab initio* results to analogous ones from DFT and TD-DFT, with the second column of Table III providing correction energy shifts for each method as a crude summary. The TD-DFT and DFT results are very similar, indicating the success of the empirical scheme Eqns. (1) – (5) and the dominant closed-shell nature of the ground state. Of the density functionals, CAM-B3LYP, which contains corrections affecting the charge-transfer contributions to the states, performs best with related differences to MRCI of $0.0\pm0.3$ eV and $0.1\pm0.2$ eV, respectively. HSE06 does not provide basic support for charge-transfer transitions [36] but nevertheless performs well for the states of interest, with an average difference of -$0.3\pm0.2$ eV compared to MRCI and -$0.3\pm0.1$ eV compared to CCSD(T). It hence has an accuracy similar to that of CCSD. PBE performs poorly in absolute terms, with related differences of -$0.8\pm0.3$ eV and -$1.0\pm0.3$ eV, respectively, but the standard deviations remain low and hence relative state orderings are maintained. Overall, these results parallel those for $V_NC_B$ [34], with the differences here much reduced in magnitude: for $V_NC_B$, only CAM-B3LYP provided a realistic description, with much larger ones reported for HSE06 and for PBE. For the low-energy excited state $(1)^2A_1$, this analysis has been extended to model compounds containing 3 rings [35], revealing that calculated excitation energies change by less than 0.1 eV, with the differences between methods preserved. Hence the analysis in Table III is expected is maintained, provided that no untoward effects manifest, e.g., as expected for charge-transfer bands determined using HSE06 and PBE [36].

E. Adiabatic transition energies and reorganization energies of $V_NN_B$.

Table IV lists the HSE06 adiabatic energies of 13 excited states of $V_NN_B$, along with the reorganization energies $\lambda_a$ and $\lambda_e$ depicting the absorption and emission spectroscopic widths, respectively, for both the model compound and the 2D material (Fig. 1). Some results are also presented using the PBE and CAM-B3LYP density functionals. Large differences are predicted between the excitation energies of the model compound and those for the 2D material of up to ± 1 eV. Two effects contribute to this: the model compound can accommodate too much motion, artificially lowering transition energies [26], and the small model compound used do not admit influences exerted by the CB and VB, increasing transition energies. The calculated reorganization energies are in much better agreement, however, indicating that, for this defect, the geometrical effect is minor compared to that arising from the involvement of orbitals from outside the defect core.

The HSE06 values for the reorganization energies of $(1)^2A_1$ and $(1)^4A_1$ in the 2D material match well with previous calculations [30]. Mostly the reorganization energies calculated for emission and absorption match, indicating the often-expected emission-absorption symmetry, but for other states the differences are large, typically indicative of the operation of large Duschinsky rotation effects that would control details of internal defect photochemical processes, as is now being commonly observed for aromatic chromophores [100].

The reorganization energies listed in Table IV include only the contributions from in-plane relaxation within $C_{2v}$, excluding any that may arise from in-plane or out-of-plane distortions. As is known [42], the $(1)^1B_1$ ground state is predicted by HSE06 to undergo a distortion in a $b_1$ mode, and we find that the same also applies to $(2)^1B_1$ by normal-mode analysis of its Hessian matrix at $C_{2v}$ symmetry.

A summary of the best-estimate spectroscopic properties of $V_NN_B$ is illustrated in Fig. 4. As inclusion of structures much larger than the single-ring model compound are critical to the properties of this defect, we start with HSE06-calculated transition energies for the 2D material. To these, we add corrections based on the calibration of HSE06 against *ab initio* methods for the model compound. Such corrections would normally be though to be both accurate and transferrable, but the energies of charge-transfer transitions in the 2D model results are likely to be underestimated [36] in a way not included in the calibration data. The $(1)^2A_1 \to (1)^2B_1$ emission consistent in energy with many observed single-photon emitters, but the calculated oscillator strength listed in Table IV is far too low to account for the observed short photoemission lifetimes, and the calculated emission reorganization energy of 0.42 eV is 4-10 times too large to account for observed spectral widths [29]. Hence these results indicate that $V_NN_B$ is not a commonly observed single-photon emitter.

### F. The +1 and -1 charged forms of $V_NN_B$.

Engineering of the Fermi energy in h-BN could, in principle, lead to the charging of the defect in either its natural environment or an artificial one. Table V shows properties of the $V_NN_B^{+1}$ and $V_NN_B^{-1}$ defect states, evaluated on the model compound using HSE06 and CAM-B3LYP, as well as on the 2D material using HSE06 in broken symmetry [101]. Both ions are predicted to have $(1)^1A_1$ ground states in both the model compound and the 2D material, arising from closed-shell electronic configurations. A $(1)^1A_1$ ground state was also previously predicted for $V_NC_B$, a system isoelectronic to $V_NN_B^{+1}$ that also can display $C_{2v}$ symmetry [28]. These closed-shell ground states depict covalent bonds forming between defect atoms, but the associated bond lengths far exceed those typical of covalent bonding (see Fig. 1). As a result, the ground state has in reality large open-shell character, meaning that single-reference computational approaches such as those applied in Table V can introduce large errors.

For $V_NC_B$, we found that CAM-B3LYP calculations appeared to give the most reliable results, but computational tools are not currently available that would allow this method to be applied to 2D materials [28]. For HSE06, we found corrections of 0.7 eV are needed for triplet states compared to closed-shell singlet states and 1.0 eV for open-shell singlet states. For the isoelectronic species $V_NN_B^{+1}$, applying these corrections bring the HSE06 and CAM-B3LYP results in Table V into line, whereas they already agree well for $V_NN_B^{-1}$ without correction. These corrections therefore do not appear universal, but rather depend on the nature of the frontier orbitals. We present tentative descriptions of the low-lying excited-state energetics of $V_NN_B^{+1}$ and $V_NN_B^{-1}$, calculated at HSE06 level and then corrected, in Fig. 5.

Table V shows that the naïve expectation that absorption and emission reorganization energies should be similar is mostly met, again vindicating the quality of the computational procedures developed. Significant systematic differences are found for the related states $(2)^1B_1$ and $(2)^3B_1$ of the cation, however, and for the important state $(1)^1B_2$ of the anion. These calculations appear robust, but are wisely treated as being suspicious.

For $V_NN_B^{+1}$, the calculations predict the lowest-energy singlet excited state to be $(1)^1B_1$. This state would have an out-of-plane transition with the ground state that is inconsistent with commonly observed h-BN defect spectra [29]. It is predicted to have very large reorganization energies ($\lambda_a, \lambda_e$ = 1.82, 1.31 eV) in the 2D material, and hence extremely broad spectra, with very low calculated oscillator strength, additional features all inconsistent with experiment. Within the triplet manifold, the low energy transitions are predicted to be either forbidden or else very weak and out-of-plane polarized, again inconsistent with experiment.

For $V_NN_B^{-1}$, the calculations again predict the lowest-energy singlet excited state to be $(1)^1B_1$, having a weak and out-of-plane polarized transition to the ground state, but this time with only a small reorganization energy of $\lambda_e$ = 0.26 eV. However, above it by just 0.5 eV are predicted to be the $(1)^1B_2$ and $(2)^1A_1$ states, states with symmetries consistent with the observed dual in-plane absorption polarizations [18]. It could be that the calculations have just misrepresented the energy of $(1)^1B_1$, and that, in reality, the other states are of lower energy. Indeed, the $(2)^1A_1$ state shows all the required properties: intense, with very small reorganization energy of just $\lambda_e$ = 0.11 eV, but $(1)^1B_2$ appears to be too weak to participate. The calculated reorganization energies of $\lambda_a$ = 0.79 eV and $\lambda_e$ = 0.27 eV appear suspicious, but if such a scenario were real then it depicts interesting properties: the second excited state is narrow in emission and so does not become the emitting state, whilst it is broad in absorption, allowing for a wide variation in excitation wavelength to drive photoluminescence. In summary, while some required features are present, overall it would seem unlikely that these states could explain the observed h-BN emission. The triplet manifold of $V_NN_B^{-1}$ is predicted to have many low energy transitions and is inconsistent with the commonly observed h-BN defect spectra [29].

G. Spin-spin interactions and possible applications to quantum information.

While none of the defects studied appear to relate to observed h-BN defects, it could be possible to engineer these defects intentionally. One of the driving forces for research into h-BN defects is the possibility that they may prove useful as qubits in quantum information processing systems. We proceed by examining how useful the calculated defects are likely to be in this regard.

Spin-orbit coupling can mix the triplet and singlet spin states, generating intersystem crossings, while spin-spin interaction lifts the degeneracy of spin multiplets. This can reveal critical properties concerning the optical cycle of a defect [102], and indeed controls [26] optically detected magnetic resonance (ODMR) observed [24] for the $V_B^-$ defect. Specific patterns of spin-orbit couplings and zero-field splittings are required for defects to be useful as qubits, and the calculated patterns for the charged defects are shown in Fig. 5. A detailed discussion on the nature of the spin-orbit interactions [103], zero-field splitting, and allowed intersystem crossing and optical transitions are complex and are provided in SM Section S3 [103-105].

The key conclusion reached is that, in subsequent optical cycles, ODMR contrast can be achieved by microwave excitation for $V_NN_B^{+1}$ and $V_NN_B^{-1}$, owing to the lifetime differences of the first and second order transitions from the different triplet sub-states to the singlet ground state. This behavior is similar to that found for the $N_2V$ defect in diamond. Thus $V_NN_B^{+1}$ and $V_NN_B^{-1}$, should be exploitable to realize a long-living quantum memory in h-BN, as has been achieved for the $N_2V$ defect in diamond [104]. Indeed these have similar electronic-state structures.

IV. CONCLUSIONS

The *a priori* calculation of properties of defect states in 2D materials remains a daunting challenge. Here, we introduce both basic theory and computing techniques to enable many computational packages to performing calculations on periodic systems to treat some rather unusual electronic states that can arise in defects. These methods are then used to look at neutral and charged states of the $V_NN_B$ defect of h-BN. The charged states have an even number of electrons, with $V_NN_B^{+1}$ being isoelectronic with a previously well-studied defect, $V_NC_B$ [34]. Both of these defects result in very large discrepancies when DFT methods are compared to *ab initio* ones for model compounds, the cause being the intrinsic open-shell nature of the ground state and other key defect states. The error here is an intrinsic one belonging to the DFT approaches used, and is not affected by use of a 2D periodic model instead of a model compound. The problems generated by the open-shell nature of key states is less for $V_NN_B^{-1}$ than it is for $V_NN_B^{+1}$, owing to the change in the nature of the key HOMO orbital. Universal corrections for DFT calculations therefore cannot be obtained; at least every isoelectronic system needs to be treated independently. In addition, we see for this defect that 1-ring model compounds show large differences in state energies compared to 2D materials, owing to the importance of transitions involving the VB and CB.

For neutral $V_NN_B$, a system with an odd number of electrons, we find here that the errors in DFT are much reduced, with state energies for model compounds requiring corrections of on average $0.0 \pm 0.3$ eV for CAM-B3LYP, $0.3 \pm 0.2$ eV for HSE06, and, as expected, large values of $0.9 \pm 0.3$ eV for PBE.

Concerning, the possibility that $V_NN_B$ defects could contribute to the commonly observed defect spectroscopy of h-BN, from the calculations performed it is clearly unlikely. Nevertheless, we note that the calculations predict that if by some means $V_NN_B$ defects could be engineered into h-BN and charged states +1 and -1 could be realized, then these defects should be useful for quantum computing applications that access different defect states. Our prediction is that $V_NN_B^{+1}$ and $V_NN_B^{-1}$ should be exploitable to realize a long-living quantum memory in h-BN, as has been achieved for $N_2V$ defect in diamond.

## ACKNOWLEDGMENTS

This work was supported by resources provided by the National Computational Infrastructure (NCI), and Pawsey Supercomputing Center with funding from the Australian Government and the Government of Western Australia, as well as Chinese NSF Grant #11674212. Computational facilities were also provided by the UTS eResearch High Performance Computer Cluster and the ICQMS Shanghai University High Performance Computer Facility. S.A. acknowledges receipt of an Australian Postgraduate Award funded by ARC DP 150103317. Funding is also acknowledged from ARC DP 160101301, as well as Shanghai High-End Foreign Expert grants to R.K. and M.J.F..

TABLE I. Considered state symmetries, their dominant orbital occupancies, and their transition polarizations from the ground state, for various charged states of the $V_N B_N$ defect in h-BN.

| | State | Trans. pol.[a] | $V_{2a1}$ | $V_{2b1}$ | $V_{1a1}$ | $1a_1$ | $V_{2a2}$ | $V_{1b1}$ | $2a_1$ | $V_{1a2}$ | $1b_1$ | $2b_1$ | $C_{1a2}$ | $C_{2a2}$ | $C_{1b1}$ | $C_{1a1}$ | $C_{2a1}$ | $C_{1b2}$ | $C_{2b1}$ | $C_{3a2}$ | Spin comp.[b] |
|---|---|---|---|---|---|---|---|---|---|---|---|---|---|---|---|---|---|---|---|---|---|
| $V_N N_B$ | $(1)^2B_1$ | - | 2 | 2 | 2 | 2 | 2 | 2 | 2 | 2 | 1 | 0 | 0 | 0 | 0 | 0 | 0 | 0 | 0 | 0 | |
| | $(2)^2B_1$ | $A_1$ | 2 | 2 | 2 | 2 | 2 | 2 | 2 | 2 | 0 | 1 | 0 | 0 | 0 | 0 | 0 | 0 | 0 | 0 | |
| | $(1)^2A_1$ | $B_1$ | 2 | 2 | 2 | 2 | 2 | 2 | 1 | 2 | 2 | 0 | 0 | 0 | 0 | 0 | 0 | 0 | 0 | 0 | |
| | $(1)^2A_2$ | $B_2$ | 2 | 2 | 2 | 2 | 2 | 2 | 2 | 2 | 0 | 0 | 1 | 0 | 0 | 0 | 0 | 0 | 0 | 0 | |
| | $(2)^2A_2$ | $B_2$ | 2 | 2 | 2 | 2 | 2 | 2 | 2 | 1 | 2 | 0 | 0 | 0 | 0 | 0 | 0 | 0 | 0 | 0 | |
| | $(2)^2A_1$ | $B_1$ | 2 | 2 | 2 | 2 | 2 | 2 | 1 | 2 | 1 | 1 | 0 | 0 | 0 | 0 | 0 | 0 | 0 | 0 | D3 |
| | $(3)^2A_2$ | $B_2$ | 2 | 2 | 2 | 2 | 2 | 2 | 2 | 1 | 1 | 1 | 0 | 0 | 0 | 0 | 0 | 0 | 0 | 0 | D3 |
| | $(4)^2A_2$ | $B_2$ | 2 | 2 | 2 | 2 | 2 | 2 | 2 | 2 | 0 | 0 | 0 | 1 | 0 | 0 | 0 | 0 | 0 | 0 | |
| | $(3)^2B_1$ | $B_1$ | 2 | 2 | 2 | 2 | 2 | 2 | 2 | 2 | 0 | 0 | 0 | 0 | 1 | 0 | 0 | 0 | 0 | 0 | |
| | $(4)^2B_1$ | $B_1$ | 2 | 2 | 2 | 2 | 2 | 2 | 2 | 2 | 0 | 0 | 0 | 0 | 0 | 0 | 0 | 0 | 1 | 0 | |
| | $(3)^2A_1$ | $A_1$ | 2 | 2 | 2 | 2 | 2 | 2 | 1 | 2 | 1 | 1 | 0 | 0 | 0 | 0 | 0 | 0 | 0 | 0 | D1 |
| | $(5)^2A_2$ | $A_2$ | 2 | 2 | 2 | 2 | 2 | 2 | 2 | 1 | 1 | 1 | 0 | 0 | 0 | 0 | 0 | 0 | 0 | 0 | D1 |
| | $(4)^2A_1$ | $A_1$ | 2 | 2 | 2 | 2 | 2 | 2 | 2 | 2 | 0 | 0 | 0 | 0 | 0 | 1 | 0 | 0 | 0 | 0 | |
| | $(6)^2A_2$ | $A_2$ | 2 | 2 | 2 | 2 | 2 | 2 | 2 | 2 | 0 | 0 | 0 | 0 | 0 | 0 | 0 | 0 | 0 | 1 | |
| | $(1)^4A_1$ | - | 2 | 2 | 2 | 2 | 2 | 2 | 1 | 2 | 1 | 1 | 0 | 0 | 0 | 0 | 0 | 0 | 0 | 0 | |
| | $(1)^4A_2$ | $A_2$ | 2 | 2 | 2 | 2 | 2 | 2 | 1 | 1 | 1 | 1 | 0 | 0 | 0 | 0 | 0 | 0 | 0 | 0 | |
| | $(1)^4B_1$ | $B_1$ | 2 | 2 | 2 | 2 | 2 | 1 | 2 | 2 | 1 | 1 | 0 | 0 | 0 | 0 | 0 | 0 | 0 | 0 | |
| | $(1)^4B_2$ | $B_2$ | 2 | 2 | 2 | 2 | 2 | 1 | 2 | 1 | 0 | 1 | 0 | 0 | 0 | 0 | 0 | 0 | 0 | 0 | |
| $V_N N_B^+$ | $(1)^1A_1$ | - | 2 | 2 | 2 | 2 | 2 | 2 | 2 | 2 | 0 | 0 | 0 | 0 | 0 | 0 | 0 | 0 | 0 | 0 | |
| | $(1)^1B_2$ | $B_2$ | 2 | 2 | 2 | 2 | 2 | 2 | 2 | 1 | 1 | 0 | 0 | 0 | 0 | 0 | 0 | 0 | 0 | 0 | S |
| | $(1)^1B_1$ | $B_1$ | 2 | 2 | 2 | 2 | 2 | 2 | 1 | 2 | 1 | 0 | 0 | 0 | 0 | 0 | 0 | 0 | 0 | 0 | S |
| | $(2)^1A_1$ | $A_1$ | 2 | 2 | 2 | 2 | 2 | 1 | 2 | 2 | 1 | 0 | 0 | 0 | 0 | 0 | 0 | 0 | 0 | 0 | S |
| | $(2)^1B_1$ | $B_1$ | 2 | 2 | 1 | 2 | 2 | 2 | 2 | 2 | 1 | 0 | 0 | 0 | 0 | 0 | 0 | 0 | 0 | 0 | S |
| | $(2)^1B_2$ | $B_2$ | 2 | 2 | 2 | 2 | 1 | 2 | 2 | 2 | 1 | 0 | 0 | 0 | 0 | 0 | 0 | 0 | 0 | 0 | S |
| | $(3)^1B_1$ | $B_1$ | 1 | 2 | 2 | 2 | 2 | 2 | 2 | 2 | 1 | 0 | 0 | 0 | 0 | 0 | 0 | 0 | 0 | 0 | S |
| | $(3)^1A_1$ | $A_1$ | 2 | 1 | 2 | 2 | 2 | 2 | 2 | 2 | 1 | 0 | 0 | 0 | 0 | 0 | 0 | 0 | 0 | 0 | S |
| | $(1)^3B_1$ | - | 2 | 2 | 2 | 2 | 2 | 2 | 1 | 2 | 1 | 0 | 0 | 0 | 0 | 0 | 0 | 0 | 0 | 0 | |
| | $(1)^3B_2$ | $A_2$ | 2 | 2 | 2 | 2 | 2 | 2 | 2 | 1 | 1 | 0 | 0 | 0 | 0 | 0 | 0 | 0 | 0 | 0 | |
| | $(1)^3A_1$ | $B_1$ | 2 | 2 | 2 | 2 | 2 | 1 | 2 | 2 | 1 | 0 | 0 | 0 | 0 | 0 | 0 | 0 | 0 | 0 | |
| | $(2)^3B_2$ | $A_2$ | 2 | 2 | 2 | 2 | 1 | 2 | 2 | 2 | 1 | 0 | 0 | 0 | 0 | 0 | 0 | 0 | 0 | 0 | |
| | $(2)^3B_1$ | $A_1$ | 2 | 2 | 2 | 1 | 2 | 2 | 2 | 2 | 1 | 0 | 0 | 0 | 0 | 0 | 0 | 0 | 0 | 0 | |
| | $(2)^3A_1$ | $B_1$ | 2 | 2 | 2 | 2 | 2 | 2 | 1 | 2 | 0 | 1 | 0 | 0 | 0 | 0 | 0 | 0 | 0 | 0 | |
| $V_N N_B^-$ | $(1)^1A_1$ | - | 2 | 2 | 2 | 2 | 2 | 2 | 2 | 2 | 2 | 0 | 0 | 0 | 0 | 0 | 0 | 0 | 0 | 0 | |
| | $(1)^1B_2$ | $B_2$ | 2 | 2 | 2 | 2 | 2 | 2 | 2 | 2 | 1 | 0 | 1 | 0 | 0 | 0 | 0 | 0 | 0 | 0 | S |
| | $(2)^1A_1$ | $A_1$ | 2 | 2 | 2 | 2 | 2 | 2 | 2 | 2 | 1 | 1 | 0 | 0 | 0 | 0 | 0 | 0 | 0 | 0 | S |
| | $(1)^1B_1$ | $B_1$ | 2 | 2 | 2 | 2 | 2 | 2 | 2 | 2 | 1 | 0 | 0 | 0 | 0 | 1 | 0 | 0 | 0 | 0 | S |
| | $(1)^1A_2$ | $A_2$ | 2 | 2 | 2 | 2 | 2 | 2 | 2 | 2 | 1 | 0 | 0 | 0 | 1 | 0 | 0 | 0 | 0 | 0 | S |
| | $(2)^1B_1$ | $B_1$ | 2 | 2 | 2 | 2 | 2 | 2 | 2 | 2 | 1 | 0 | 0 | 0 | 0 | 0 | 1 | 0 | 0 | 0 | S |
| | $(2)^1B_2$ | $B_2$ | 2 | 2 | 2 | 2 | 2 | 2 | 2 | 2 | 1 | 0 | 0 | 1 | 0 | 0 | 0 | 0 | 0 | 0 | S |
| | $(1)^3A_1$ | - | 2 | 2 | 2 | 2 | 2 | 2 | 2 | 2 | 1 | 1 | 0 | 0 | 0 | 0 | 0 | 0 | 0 | 0 | |
| | $(1)^3B_2$ | $B_2$ | 2 | 2 | 2 | 2 | 2 | 2 | 2 | 2 | 1 | 0 | 1 | 0 | 0 | 0 | 0 | 0 | 0 | 0 | |
| | $(1)^3B_1$ | $B_1$ | 2 | 2 | 2 | 2 | 2 | 2 | 2 | 2 | 1 | 0 | 0 | 0 | 0 | 1 | 0 | 0 | 0 | 0 | |
| | $(2)^3B_2$ | $B_2$ | 2 | 2 | 2 | 2 | 2 | 2 | 2 | 2 | 1 | 0 | 0 | 1 | 0 | 0 | 0 | 0 | 0 | 0 | |
| | $(2)^3A_1$ | $A_1$ | 2 | 2 | 2 | 2 | 2 | 2 | 2 | 2 | 1 | 0 | 0 | 0 | 1 | 0 | 0 | 0 | 0 | 0 | |
| | $(1)^3A_2$ | $A_2$ | 2 | 2 | 2 | 2 | 2 | 2 | 2 | 2 | 1 | 0 | 0 | 0 | 0 | 0 | 0 | 1 | 0 | 0 | |

a: Franck-Condon allowed $A_1$ transitions are polarized in-plane along the symmetry axis, $A_2$ transitions are forbidden, $B_1$ transitions are polarized perpendicular to the layer, while $B_2$ transitions are polarized in plane and orthogonal to the symmetry axis. Transition symmetries to/from the lowest-energy state of each spin multiplicity and ionization level are listed.

b: Spin components for open-shell calculations involving multiple degenerate spin components: S- singlet energy from Eqn. (2), D3 and D1- tripdoublet and singdoublet energies from Eqn. (5).

TABLE II. Energies (in eV) of various excited states of the single-ring $V_NN_B$ model compound (Fig. 1a) with respect to $(1)^2B_1$, performed at a reference geometry determined for $(1)^2B_1$, depicting calculated vertical excitation energies.

| State | MRCI | CASPT2 | CCSD (T) | CCSD | EOM-CCSD | TD CAM-B3LYP | TD HSE06 | TD PBE | HSE06 G09 | PBE VASP | HSE06 VASP |
|---|---|---|---|---|---|---|---|---|---|---|---|
| $(2)^2B_1$ | | 3.35 | | | 3.99 | 3.69 | 3.04 | 2.23 | | 2.46 | 3.38 |
| $(1)^2A_1$ | 3.19 | 3.50 | 3.26 | 3.35 | 3.34 | 3.18 | 3.00 | 2.47 | 2.96 | 2.54 | 2.93 |
| $(1)^2A_2$ | 3.61 | 3.15 | 3.15 | 3.43 | 3.57 | 3.48 | 3.10 | 2.38 | 3.06 | 2.55 | 3.00 |
| $(2)^2A_2$ | 3.81 | 3.89 | 4.21 | 4.88 | 4.59 | 4.34 | 3.84 | 2.74 | 3.87 | 3.10 | 3.60 |
| $(2)^2A_1$ | 4.37 | 4.10 | | | 4.48 | 4.16 | 4.82 | 3.99 | | | |
| $(3)^2A_2$ | | 4.31 | | | 4.98 | 4.78 | 4.46 | 4.19 | | | |
| $(4)^2A_2$ | | 4.58 | | | 5.21 | 5.05 | 4.61 | 3.84 | 4.42 | 3.74 | 4.39 |
| $(1)^4A_1$ | 4.36 | 4.20 | 4.38 | 4.48 | | 4.28 | 3.89 | 3.52 | 3.94 | 3.55 | 3.88 |
| $(3)^2B_1$ | | 5.03 | | | 5.57 | 5.31 | 4.76 | 3.87 | | 4.17 | |
| $(3)^2A_1$ | | 4.95 | 5.25 | 5.54 | 5.43 | 5.12 | 5.32 | 4.18 | | 4.48 | 5.17 |
| $(5)^2A_2$ | | 4.83 | | | 5.52 | 5.35 | 5.06 | 4.29 | | | |
| $(4)^2A_1$ | | 5.27 | | | 5.57 | 5.63 | 5.21 | 4.36 | | | 4.59 |
| $(6)^2A_2$ | | 4.52 | | | | 6.25 | 5.31 | 5.53 | | | |

TABLE III. Corrections to add to DFT calculated transition energies from $(1)^2B_1$ (second column, in eV), determined from averaged differences in transition energies predicted by the methods listed in the rows compared to those in the later columns, for the $V_NN_B$ model compound (Fig. 1b), evaluated at a test geometry.[a]

| method | Correction | Difference analyses (number of comparisons, average difference ± standard deviation) | | | | | | | |
|---|---|---|---|---|---|---|---|---|---|
| | | MRCI (13,13) | CCSD(T) | CCSD | EOM-CCSD | CASPT2 | TD CAM-B3LYP | HSE06 | TD PBE |
| CCSD(T) | | 4<br>0.0±0.3 | | | | | | | |
| CCSD | | 4<br>0.3±0.5 | 5<br>0.3±0.2 | | | | | | |
| EOM-CCSD | | 4<br>0.3±0.3 | 4<br>0.3±0.1 | 4<br>-0.0±0.2 | | | | | |
| CASPT2 | | 5<br>-0.1±0.3 | 5<br>-0.0±0.2 | 5<br>-0.3±0.4 | 11<br>-0.5±0.2 | | | | |
| TD CAM-B3LYP | 0.0±0.3 | 5<br>0.0±0.3 | 5<br>0.1±0.2 | 5<br>-0.2±0.2 | 11<br>-0.2±0.1 | 13<br>0.4±0.4 | | | |
| HSE06 | -0.3±0.2 | 4<br>-0.3±0.2 | 4<br>-0.3±0.1 | 4<br>-0.6±0.3 | 3<br>-0.5±0.1 | 4<br>-0.2±0.2 | 4<br>-0.4±0.1 | | |
| TD HSE06 | -0.2±0.4 | 5<br>-0.1±0.4 | 5<br>-0.2±0.2 | 5<br>-0.5±0.3 | 11<br>-0.5±0.3 | 13<br>0.1±0.4 | 13<br>-0.3±0.4 | 4<br>0.0±0.0 | |
| TD PBE | -0.9±0.3 | 5<br>-0.8±0.3 | 5<br>-1.0±0.3 | 5<br>-1.2±0.5 | 11<br>-1.2±0.4 | 13 -0.6±0.6 | 13<br>-1.0±0.4 | 4<br>-0.7±0.3 | 13<br>-0.7±0.4 |

[a] calculations are performed at a geometry optimized for the $(1)^2B_1$ reference state, see text.

TABLE IV. Energies with respect to the $(1)^1B_1$ ground state of various excited states of the $V_NN_B$ model compound and 2-D material (Fig. 1), evaluated at their adiabatic minimum-energy geometries constrained to $C_{2v}$ symmetry (see SM Section S5), oscillator strengths in absorption calculated using TDDFT, and the reorganization energies $\lambda_a$ and $\lambda_e$ depicting the width of absorption and emission bands, respectively.

| State | Adiabatic excitation energy / eV | | | | | | Osc. Strength | | $\lambda_a$ / eV | | | | | | $\lambda_e$ / eV |
|---|---|---|---|---|---|---|---|---|---|---|---|---|---|---|---|
| | Model | | | | 2D material | | model | | model | | | | 2D material | | 2D material |
| | PBE | HSE06 G09 | HSE06 VASP | TD CAM | PBE | HSE06 | HSE06 | CAM | PBE | HSE06 G09 | HSE06 VASP | TD CAM | PBE | HSE06 | HSE06 |
| $(2)^2B_1$ | 2.14 | | 2.86 | 2.37 | 2.04 | 2.39 | 0.0877 | 0.0985 | 0.31 | | 0.52 | 0.54 | 0.49 | 0.73 | 0.72 |
| $(1)^2A_1$ | 2.03 | 2.44 | 2.41 | 1.93 | 1.29 | 1.81 | 0.0001 | 0.0002 | 0.52 | 0.53 | 0.52 | 0.48 | 0.60 | 0.57 | 0.42 |
| $(1)^2A_2$ | 2.17 | 2.50 | 2.51 | 2.23 | 2.80 | 3.36 | 0.0023 | 0.0019 | 0.38 | 0.56 | 0.51 | 0.47 | 0.32 | 0.45 | 0.44 |
| $(2)^2A_2$ | 2.84 | 3.55 | 3.27 | 2.25 | 2.85 | 3.36 | 0.0003 | 0.0007 | 0.26 | 0.32 | 0.33 | 0.15 | 0.21 | 0.45 | 0.44 |
| $(2)^2A_1$ | 3.53 | | 4.60 | 3.63 | 2.70 | 3.48 | 0.0001 | 0.0002 | | | | 0.25 | | | |
| $(3)^2A_2$ | 4.03 | | 4.68 | 4.12 | 3.98 | 4.76 | 0.0126 | 0.0066 | | | | 0.12 | | | |
| $(4)^2A_2$ | 3.43 | 4.06 | 4.00 | 3.70 | 3.02 | | 0.0564 | 0.0639 | 0.31 | 0.36 | 0.39 | 0.57 | 0.39 | | |
| $(1)^4A_1$ | 3.27 | 3.60 | 3.54 | 2.76 | 2.46 | 2.69 | 0 | 0 | 0.28 | 0.34 | 0.34 | | 0.29 | 0.26 | 0.40 |
| $(3)^2B_1$ | 3.80 | | 4.47 | 4.40 | 2.99 | 4.88 | 0.0121 | 0.0388 | 0.37 | | 0.19 | 0.13 | 0.36 | 0.16 | 0.41 |
| $(3)^2A_1$ | 4.19 | | 5.10 | 4.28 | 3.36 | 3.98 | 0.0066 | 0.0085 | | | | 0.06 | | | |
| $(5)^2A_2$ | 4.13 | | 5.28 | 4.99 | 4.08 | 5.36 | 0.0460 | 0.0601 | | | | 0.42 | | | |
| $(4)^2A_1$ | 3.72 | | 4.56 | | 2.96 | 3.56 | 0.0006 | 0.0012 | | | 0.03 | | | | |

TABLE V. Energies of various excited states of the $V_NN_B^{+1}$ and $V_NN_B^{-1}$ model compound and related periodic 2-D material (Fig. 1), with respect to their $(1)^1A_1$ ground states, evaluated at their adiabatic minimum-energy geometries constrained to $C_{2v}$ symmetry (see SM Section S5), oscillator strengths in absorption, and the reorganization energies[a] $\lambda_a$ and $\lambda_e$ depicting the width of absorption and emission bands, respectively.

| Defect | State | Adiabatic excitation energy / eV | | | Osc. Strength | | $\lambda_a$ / eV | | $\lambda_a$ / eV | $\lambda_e$ / eV |
|---|---|---|---|---|---|---|---|---|---|---|
| | | model | | 2D material | model TD | | model | | 2D material | 2D material |
| | | HSE06 TD | CAM TD | HSE06 | CAM | HSE06 | HSE06 TD | CAM TD | HSE06 | HSE06 |
| | $(1)^1B_2$ | 0.72 | 1.84 | 2.24 | 0.0014 | 0.0006 | 0.81 | 0.83 | 0.57 | 0.58 |
| | $(1)^1B_1$ | 1.9 | 3.27 | 0.86 | 0 | 0 | 0.82 | 0.53 | 1.82 | 1.31 |
| | $(2)^1A_1$ | 2.68 | 3.35 | 2.44 | 0.0886 | 0.0552 | 0.60 | 0.68 | 0.48 | 0.66 |
| | $(2)^1B_1$ | 3.33 | 3.6 | 3.70 | 0.0046 | 0.0029 | 0.21 | 0.26 | 0.32 | 0.59 |
| | $(2)^1B_2$ | 3.11 | 3.88 | 2.75 | 0.0454 | 0.0116 | 0.36 | 0.35 | 0.54 | 0.58 |
| | $(3)^1B_1$ | 4.00 | 4.42 | 3.79 | 0.0057 | 0 | 0.19 | 0.20 | 0.73 | 0.81 |
| $V_NN_B^{+1}$ | $(3)^1A_1$ | | | 2.30 | | | | | 0.62 | 0.61 |
| | $(1)^3B_1$ | 1.55 | 2.13 | 0.60 | - | - | - | 0.31 | - | - |
| | $(1)^3B_2$ | 1.09 | 1.96 | 2.20 | 0 | 0 | 0.66 | 0.77 | 1.28 | 1.42 |
| | $(1)^3A_1$ | 2.24 | 2.99 | 2.33 | 0.0000 | 0.0000 | 0.75 | 0.93 | 1.31 | 1.17 |
| | $(2)^3B_2$ | 2.71 | 3.48 | 2.76 | 0.0000 | 0.0000 | 0.29 | 0.28 | 1.36 | 1.08 |
| | $(2)^3B_1$ | 2.770 | 32.17 | 3.78 | 0.1038 | 0.0760 | 0.54 | 0.75 | 0.50 | 1.01 |
| | $(2)^3A_1$ | 2.93 | 3.96 | 2.78 | 0.0001 | 0.0001 | 0.45 | 0.55 | 1.24 | 1.13 |
| $V_NN_B^{-1}$ | $(1)^1B_2$ | 1.27 | 1.53 | 1.66 | 0.0001 | 0.0014 | 0.74 | 0.73 | 0.79 | 0.27 |
| | $(2)^1A_1$ | 2.13 | 2.24 | 1.66 | 0.2766 | 0.2689 | 0.27 | 0.31 | 0.11 | 0.11 |
| | $(1)^1B_1$ | 2.73 | 2.76 | 1.10 | 0.0011 | 0.0008 | 0.21 | 0.22 | 0.27 | 0.26 |
| | $(1)^1A_2$ | 2.8 | 3.23 | 1.71 | 0 | 0 | 0.42 | 0.48 | 0.26 | 0.28 |
| | $(2)^1B_1$ | 2.93 | 3.44 | 1.52 | 0.0006 | 0.0001 | 0.37 | 0.39 | 0.29 | 0.26 |
| | $(2)^1B_2$ | 3.29 | 3.39 | collapsed | 0.2237 | 0.2070 | 0.32 | 0.35 | | |
| | $(1)^3A_1$ | 1.67 | 1.11 | 0.76 | - | - | - | - | - | - |
| | $(1)^3B_2$ | 1.48 | 1.06 | 1.89 | 0.0095 | 0.0018 | 0.33 | 0.69 | 0.52 | 0.39 |
| | $(1)^3B_1$ | | -0.13 | 1.10 | 0 | 0.0001 | | 0.85 | 0.39 | 0.36 |
| | $(2)^3B_2$ | 3.08 | 2.66 | 1.98 | 0.0347 | 0.0364 | 0.21 | 0.15 | 0.47 | 0.34 |
| | $(2)^3A_1$ | 3.59 | 3.08 | 1.77 | 0.1306 | 0.1515 | 0.15 | 0.17 | 0.31 | 0.37 |
| | $(2)^3B_1$ | | | 1.50 | | | | | 0.39 | 0.43 |
| | $(1)^3A_2$ | 3.46 | 2.99 | 1.71 | 0 | 0 | 0.22 | 0.21 | 0.34 | 0.30 |

a: for relaxation from/to the $(1)^1A_1$ ground state for all singlet states and for the lowest triplet state, $(1)^3B_1$ for $V_NN_B^{+1}$ and $(1)^3A_1$ for $V_NN_B^{-1}$, else for excited triplet states for relaxation from/to the lowest triplet state, as is appropriate for triplet to triplet spectroscopy.

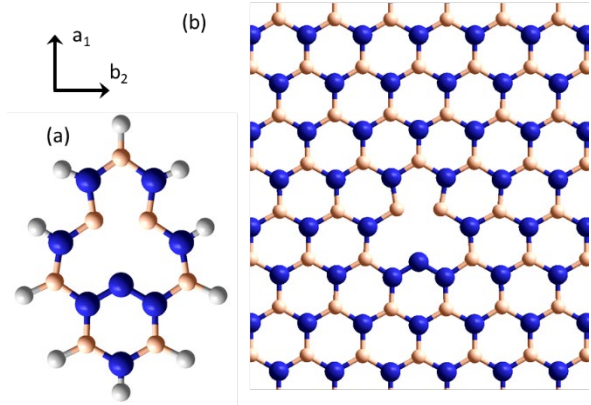

FIG. 1. The geometrical structure of the V$_N$N$_B$ defect in $C_{2v}$ symmetry, in which a nitrogen vacancy is neighbored by a nitrogen substituting boron, is represented as either (a) a model compound or (b) a rectangular (6 x 4√3)R30° unit cell representing a periodic 2D lattice of defects; N- blue, B- peach, H- white. Allowed in-plane spectroscopic transitions may be polarized along the indicated a$_1$ and b$_2$ axes, while b$_1$ transitions are polarized out-of-plane and a$_2$ transitions are forbidden.

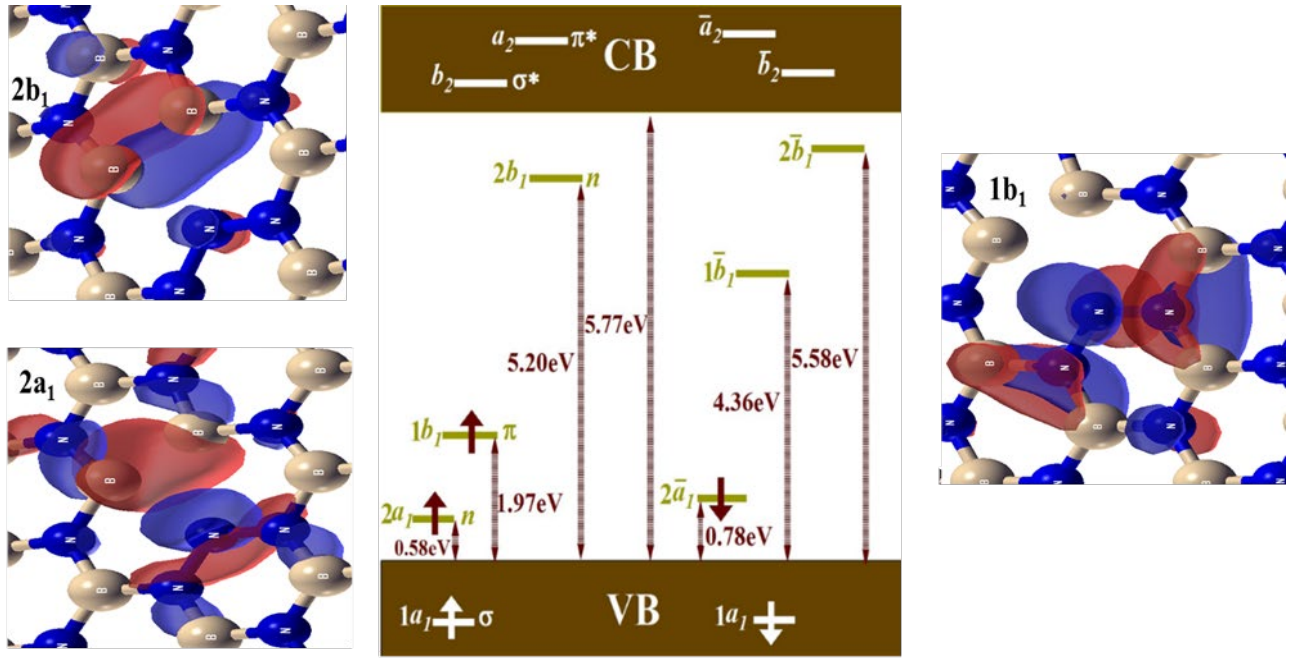

FIG. 2. Shown for the example of neutral V$_N$N$_B$ in 2D h-BN constrained to $C_{2v}$ symmetry are: (central) defect HSE06 orbitals localized within the h-BN valence-conduction band gap from $(1)^2B_1$ ground-state electronic structure, and (flanks) wavefunctions pertaining to some of the key orbitals.

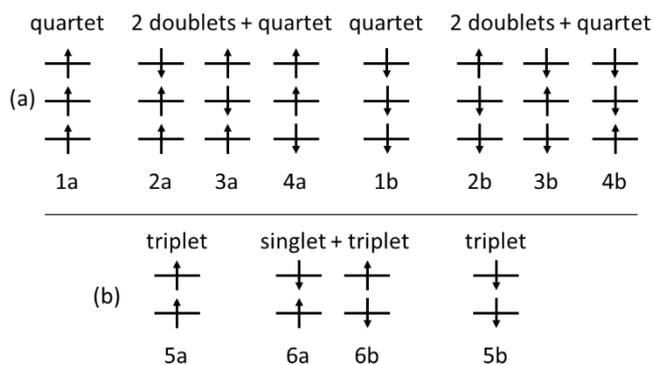

FIG. 3. Possible occupancies (a) when 3 unpaired electrons distribute into 3 orbitals (resulting in two doublet states the tripdoublet, D3, and singdoublet, D1, and one quartet), and (b) when 2 unpaired electrons distribute into 2 orbitals (resulting in one singlet state and one triplet).

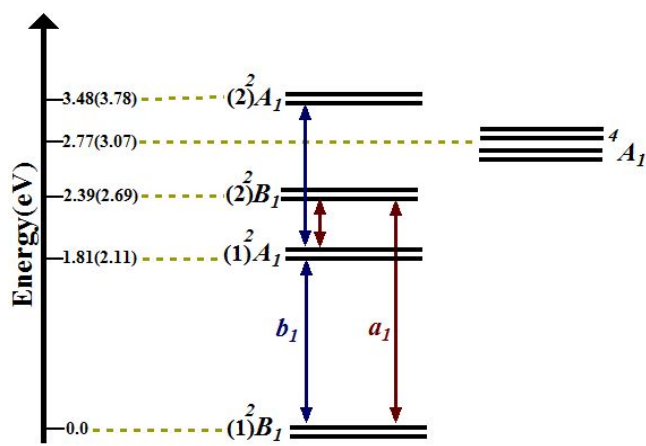

FIG 4. HSE06 adiabatic energies of low lying states of 2D $V_NN_B$ as calculated by DFT, with, in (), these energies corrected according to *ab initio* CCSD(T), EOM-CCSD, and CASPT2 calculations for the model compound. Allowed transition polarizations are also indicated.

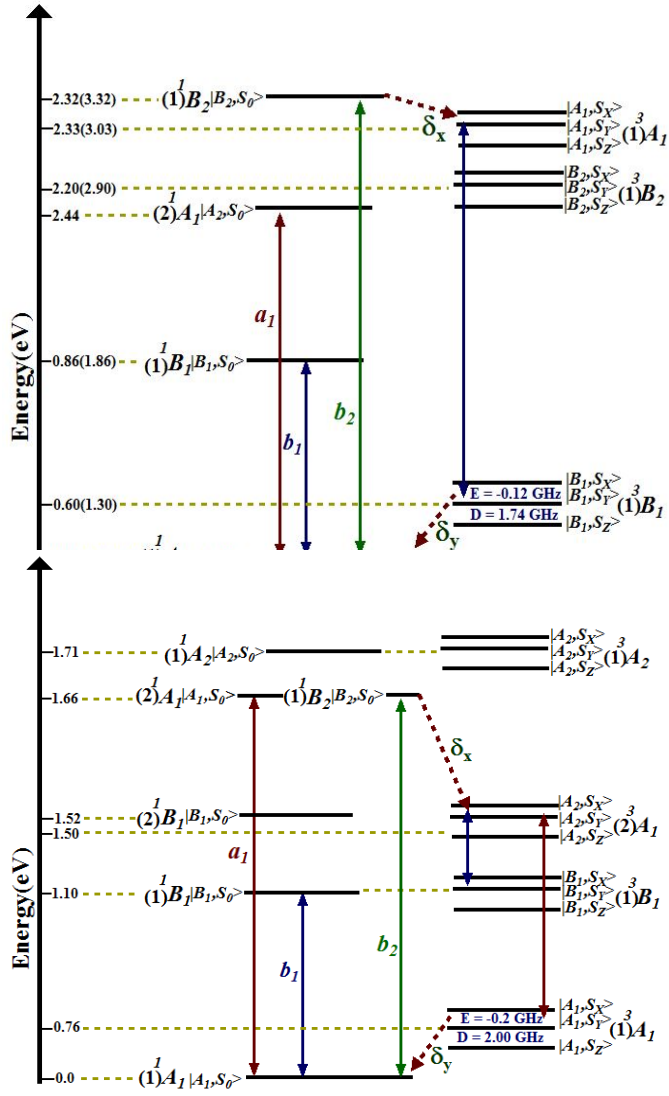

FIG 5. HSE06 adiabatic energies of low lying states of 2D $V_NN_B^{+1}$ (left) and $V_NN_B^{-1}$ (right), as calculated by HSE06, with, in () for the anion, these energies corrected according to *ab initio* CCSD(T), EOM-CCSD, and CASPT2 calculations for the isoelectronic defect $V_NC_B$. Allowed transition polarizations in the $a_1$, $b_1$, and $b_2$ directions (see Fig. 1) are also indicated.